# Nanoporous graphene: a 2D semiconductor with anisotropic mechanical, optical and thermal conduction properties

Bohayra Mortazavi[*,b,c], Mohamed E. Madjet[τ,d], Masoud Shahrokhi[τ,e], Said Ahzi[d], Xiaoying Zhuang[c] and Timon Rabczuk[a]

[a]*Institute of Research & Development, Duy Tan University, Quang Trung, Da Nang, Vietnam.*
[b]*Institute of Structural Mechanics, Bauhaus-Universität Weimar, Marienstr. 15, D-99423 Weimar, Germany.*
[c]*Cluster of Excellence PhoenixD (Photonics, Optics, and Engineering–Innovation Across Disciplines), Leibniz Universität Hannover, Hannover, Germany.*
[d]*Qatar Environment and Energy Research Institute, Hamad Bin Khalifa University, Doha, Qatar.*
[e]*Department of Physics, Faculty of Science, Razi University, Kermanshah, Iran.*

**Abstract**

Nanoporous graphene (NPG), consisting of ordered arrays of nanopores separated by graphene nanoribbons was recently realized using a bottom-up synthesis method (*Science 360(2018), 199*). In this work we accordingly explored the mechanical response, thermal conductivity and electronic/optical properties of single-layer NPG using the density functional theory and molecular dynamics simulations. Along the armchair direction, NPG was found to exhibit higher tensile strength and thermal conductivity by factors of 1.6 and 2.3, respectively, in comparison with the zigzag direction. Despite of showing high rigidity and tensile strength, NPG was predicted to show around two orders of magnitude suppressed thermal conductivity than graphene. Results based on GGA/PBE highlight that NPG monolayer presents semiconducting electronic character with a direct band-gap of 0.68 eV. According to the HSE06 estimation, NPG monolayer shows a band-gap of 0.88 eV, very promising for the application in nanoelectronics. Optical results reveal that NPG nanomembranes can absorb the visible, IR and NIR light. This work highlights the outstanding physics of NPG, as a novel porous carbon based two-dimensional material, which may serve as a promising candidate to design advanced nanoelectronics, nanooptics and energy conversion systems.

Corresponding authors: *bohayra.mortazavi@gmail.com, #timon.rabczuk@uni-weimar.de;
[τ]these authors contributed equally to this work

# 1. Introduction

Graphene [1,2], the most prominent member of the two-dimensional (2D) materials family show uniquely high mechanical properties [3] and thermal conductivity [4,5] and very attractive optical/electronic properties [6–9]. Outstanding properties of graphene, along with its broad-range application prospects, motivated the design and synthesis of different classes of 2D materials. Although graphene exhibits exceptional properties, it also shows few limitations for particular applications. Graphene limitations interestingly have been acting positively, and serve as motivations for the experimental and theoretical endeavours to search, predict, design and fabricate novel 2D materials during the last decade [9–12]. In this regard, pristine graphene is not a semiconductor and shows zero band-gap, which limits its application in nanoelectronics. This issue has encouraged the synthesis of a wide-variety of 2D semiconductors, such as: transition metal dichalcogenides [13–15], phosphorene [16,17] and carbon nitride [18–23] nanosheets. As another example, densely packed atomic lattice of graphene, makes the access to the active sites difficult and moreover full $sp^2$ carbon atomic type results in moderate charge capacities for the application in energy storage systems, like batteries. In this case, full carbon 2D materials with porous atomic structures like graphdiyne [24,25] can exhibit considerably higher charge capacities and provide faster diffusion rates in comparison with graphene, highly desirable for the design of next generation Li-ion battery technologies [26–29].

One of the most appealing points of graphene lies in its ability to exhibit vastly tuneable electronic, optical and thermal properties, by mechanical straining [30–34], defect engineering [35–39] or chemical doping [40–44]. In particular, creating various configurations of patterned cuts in graphene using the lithography techniques, such as the graphene kirigami [45], graphene nanomesh [46] or graphene antidot lattices [47,48] and porous graphene [49], have been successfully employed to open a band-gap in graphene [50–52], enhance its efficiency for various chemical processes [53–55] and prepare it for advanced technologies, like the DNA sequencing [56–58]. Nevertheless, the fabrication of porous graphene nanomembranes through the employment of lithography methods requires additional processing steps after the synthesis of pristine graphene, which are not only complicated in practice but also time consuming and expensive as well. This way, a more practical approach is to design a chemical process that can directly lead to the growth of porous graphene. To address this experimental challenge, very recently Moreno *et al.*

[59] reported the first successful synthesis of chemically grown nanoporous graphene (NPG) sheets using a bottom-up method. The NPG was found to exhibit inherent semiconducting electronic character and was predicted to offer desirable properties for simultaneous sieving and electrical sensing of molecular species [59]. This latest experimental success, raise the importance of theoretical and experimental investigations in order to provide more in-depth understanding about various intrinsic material properties of NPG nanosheets. In this regard, analysis of structural stability, mechanical responses, thermal conductivity, optical and electronic properties of NPG are highly essential to gain basic understanding about this novel nanosheet and finding promising application prospects. In this work our objective is to efficiently explore the aforementioned properties of single-layer and free-standing NPG by conducting atomistic simulations. We employed density functional theory based simulations to explore the mechanical, optical and electronic properties and in addition the classical molecular dynamics simulations were conducted to evaluate the phononic thermal conductivity of NPG. We hope that our results can provide general and useful vision concerning the critical properties of semiconducting NPG nanosheets and may serve as guide for the future theoretical and experimental studies.

## 2. Computational methods

The first-principles density functional theory (DFT) calculations in this work were carried out using the *Vienna Ab-initio Simulation Package* (VASP) [60–62]. The DFT calculations were conducted within the frame work of generalized gradient approximation (GGA) in the form of Perdew–Burke–Ernzerhof (PBE) [63] for the exchange correlation potential and the ion–electron interaction was treated using the projector augmented wave (PAW) [64] method. A plane-wave cutoff energy of 500 eV was used and the convergence criterion for the electronic self consistence-loop was set to be $10^{-4}$ eV. Periodic boundary conditions were applied along the all directions, considering a vacuum thickness of 15 Å along the sheet normal direction. VESTA [65] package was used for the illustration of atomic structures and charge densities as well. Energy minimized structures were acquired by altering the size of the unit-cells and then employing the conjugate gradient method for the geometry optimizations. The convergence criteria for the Hellmann–Feynman forces on each atom was taken to be 0.01 eV/Å using a 2×4×1 Monkhorst-Pack [66] k-point mesh size. Mechanical properties were evaluated by performing uniaxial tensile simulations. Because of the underestimation of band-gap by the PBE/GGA method, we used the screened hybrid functional HSE06 [67] to provide more accurate estimations. For the Brillouin zone integration, 6×16×1 and 3×5×1 Monkhorst-Pack [66] k-point meshes were used to report the electronic results within the PBE and HSE06 methods, respectively. The optical properties were evaluated using the random phase approximation (RPA) [68] constructed over the PBE method. In this case, Brillouin zone was sampled by 12×12×1 grids. A detailed explanation of optical property calculations were described in our previous works [69–71].

We conducted the non-equilibrium molecular dynamics (NEMD) simulations to predict the lattice thermal conductivity of NPG monolayers. The NEMD simulations were conducted using the LAMMPS [72] package along with the AIREBO [73] force-field to introduce the atomic interactions. In the all NEMD simulations, we applied periodic boundary condition along the planar directions and a relatively small time increment of 0.25 fs was used to count for the high vibrations of H atoms [74–78]. The NEMD calculations were performed for NPG monolayers with different lengths to investigate the length effect on the predicted thermal conductivities. To simulate the heat transfer, we first relax the structures at the room temperature using the Nosé-Hoover barostat and thermostat method (NPT). We then fixed atoms at the two ends of the sample and divided the simulation box (excluding the fixed atoms) along the heat transfer direction into 20 slabs. The first two slabs at the two ends were assigned to be cold and hot slabs, respectively. To simulate steady-state heat transfer, we applied a 20 K temperature difference between the hot and cold slabs. During this step, the 20 K temperature difference was preserved in the aforementioned slabs employing the Nosé-Hoover thermostat (NVT) method, while the rest of the system was simulated using the constant energy method (without a thermostat being applied). As a result of applied temperature difference, a steady-state heat-flux will be imposed and simultaneously a temperature gradient will establish along the heat transfer direction. To decrease the thermal fluctuations, the simulations were conducted for 7.5 ns and the temperatures at every slab and heat-flux values were averaged. The final thermal conductivity along the loading direction, $\kappa_x$, was then evaluated using the established temperature gradient *(dT/dx)* and the imposed heat-flux, $q_x$, using the one-dimensional form of the Fourier law; $\kappa_x = q_x \frac{dx}{dT}$, assuming a thickness of 3.35 Å for single-layer NPG, the same as that of the graphene.

## 3. Results and discussions

In Fig. 1, the atomic structure of energy minimized NPG monolayer is illustrated. In an analogy to graphene, one can also define two major directions of armchair and zigzag in NPG, as distinguished in Fig. 1. According to our results, the lattice constants along the zigzag and armchair directions were found to be 32.383 Å and 8.534 Å, respectively. To briefly analyse the bonding nature in NPG nanosheets, the electron localization function (ELF) [79] within the unit-cells is also illustrated in Fig. 1. ELF is a spatial function and takes a value between 0 and 1. As expected, electron localization occurs around the center of all C-C bonds in the NPG, confirming the covalent bonding. To facilitate the future studies, energy minimized unit-cell of NPG is provided in the data section.

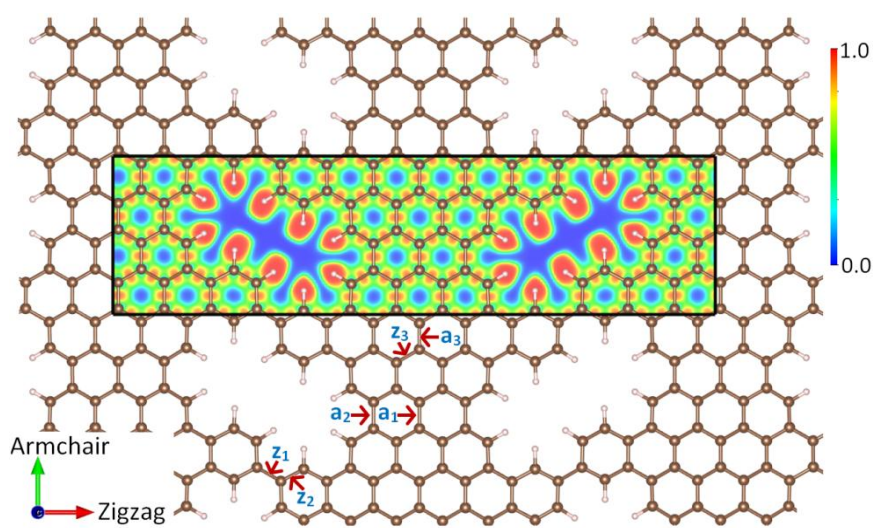

**Fig. 1**, Atomic structure of nanoporous graphene (NPG). Contours illustrate the electron localization function [79] within the unit-cell. To investigate the evolution of bond lengths during the uniaxial loading (results are shown in Fig. 2), as shown here six different bonds were chosen; three along the armchair direction ($a_1$, $a_2$ and $a_3$) and three oriented along the zigzag direction ($z_1$, $z_2$ and $z_3$).

For the application of a material in various devices, presenting good mechanical properties are among the most important requirements. This way we first study the mechanical responses of NPG nanosheet by conducting uniaxial tensile simulations. These simulations were carried out along the armchair and zigzag directions, in order to probe the anisotropy in the mechanical properties. For the uniaxial tensile simulations, the periodic simulation box size along the loading direction was increased gradually. In this case, in order to observe the uniaxial stress-conditions, the simulation box size along the sheet perpendicular direction of loading was adjusted to reach a negligible stress. After the adjustments of the simulation box sizes and accordingly rescaling the atomic positions, an energy minimization step within the conjugate gradient method was conducted to allow the rearrangement of atomic positions.

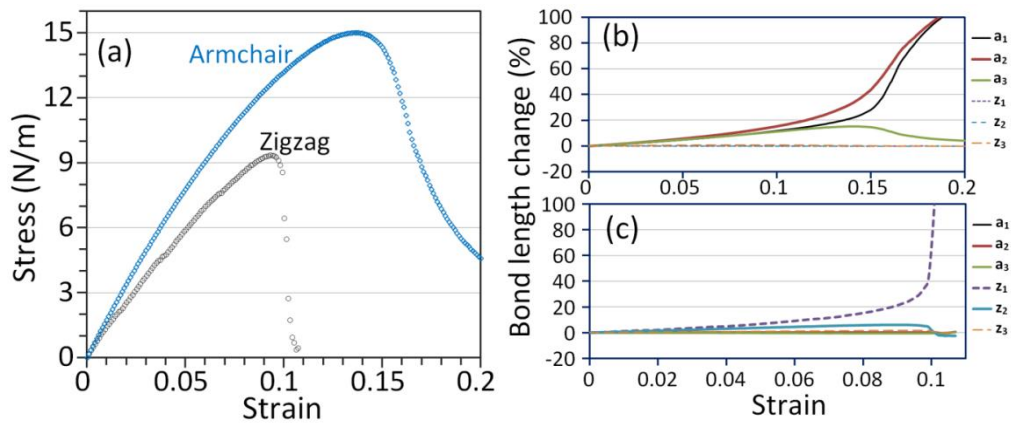

**Fig. 2**, (a) Uniaxial stress-strain responses of single-layer nanoporous graphene stretched along the armchair and zigzag directions. The evolution of some bond lengths (distinguished in Fig. 1) during the uniaxial loading along the (b) armchair and (c) zigzag directions.

In Fig. 2a, the first-principles results for the uniaxial stress-strain responses of NPG nanosheet elongated along the armchair and zigzag directions are compared. Notably, despite of the porous structure of NPG nanosheets, its uniaxial stress-strain responses along the both considered loading direction show initial linear relations, corresponding to the linear elasticity. Exhibiting linear elasticity reveals that NPG behaves analogous to densely packed and defect-free 2D materials, like; graphene, $MoS_2$ and borophene, in which the deformation is achieved mainly by the bond-elongation. We remind that in the case of graphene kirigami, at the starting points of the stretching the deformation proceeds mostly by the structural deflection rather than the bond-elongation, which can results in an order of magnitude higher stretchability than pristine graphene [52]. Observation of linear elasticity in NPG sheet suggest that its stretchability may not reach high values since the stretching of C-C bonds are limited. The results shown in Fig. 2a also reveal that initial linear and subsequent non-linear sections of the stress-strain curves are distinctly different for the uniaxial loading along the armchair and zigzag directions, confirming the anisotropic mechanical responses of NPG. To estimate the elastic modulus, linear lines were fitted to the uniaxial stress-strain values for the strain levels below 0.01. Based on our first-principles results, the elastic modulus of NPG sheet along the armchair and zigzag directions were measured to be 174 N/m and 144 N/m, respectively. The Poisson's ratio was however found to be 0.17 and convincingly independent of the loading direction. The tensile strength along the armchair and zigzag directions were found to be 15.0 N/m and 9.4 N/m, respectively. Interestingly, after reaching the ultimate tensile point, for the loading along the zigzag direction the stress values suddenly drop, whereas along the armchair direction the drop in the stress-strain curve happens very smoothly. The patterns of drops in the stress values after the tensile strength point suggest higher brittleness of NPG for the loading along the zigzag than armchair direction. Based on these findings it can be concluded that along the armchair direction, NPG exhibits higher rigidity, tensile strength, ductility and stretchability in comparison with the zigzag direction.

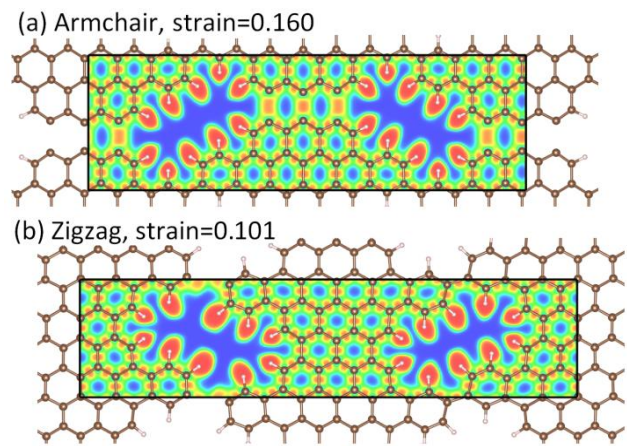

**Fig. 3**, Atomic structures of nanoporous graphene at the strain levels close to the ultimate tensile strength points for the uniaxial loading along the (a) armchair and (b) zigzag directions.

The anisotropic mechanical responses of NPG was quite expectable, because of the fact that along the zigzag direction two graphene nanoribbons are connected by a single C-C bond, whereas along the armchair in the narrowest section of the nanoribbon includes three C-C bonds. To more accurately analyze the underlying deformation mechanism, we plotted the evolution of bond lengths during the loading condition, as depicted in Fig. 2. In this case, we considered six different bonds as distinguished in Fig. 1. We first consider the uniaxial loading along the armchair direction as shown in Fig. 2b. For this loading direction, from the early stages of the loading the outer C-C bonds of the narrowest section of the nanoribbons ($a_2$ in Fig. 1) elongate more as compared with the inner C-C bond ($a_1$ in Fig. 1), which can be explained due to the higher stress-concentration around the pores tip. Interestingly, during the whole loading process, the lengths of the bonds originally oriented oblique to the zigzag direction evolve negligibly, meaning that they basically do not contribute in the load bearing. Results shown in Fig. 2b reveal that at the tensile strength point, two different changes occur simultaneously. First, the bonds oriented along the armchair direction, except those three inside the narrowest section, start to contract which results in an overall decrease in the effective stress of the system (consider $a_3$ bond in Fig. 2b). Second, the all three bonds in the narrowest section of nanoribbons keep stretching, however, the rate of stretching for the two outer bonds slow done while the C-C bond inside starts to elongate more considerably. In Fig. 3a, we plotted the NPG sample along with its corresponding ELF contour at a strain level after the tensile strength point, for the uniaxial loading along the armchair direction. From the graphical representation of the bond in this system (<2.1 Å), one may conclude that for this sample the rupture has already occurred for the all the three bonds forming the narrowest section of the graphene nanoribbons. However, the ELF contour in accordance with our bond length analysis, suggests that these aforementioned bonds are yet actively involved in the load bearing, since the ELF values are over 0.75 around the center of these bond, revealing the intactness of covalent bonding. In contrast for the loading along the zigzag direction, the single bond connecting two separate graphene nanoribbons mainly contribute to the load bearing (consider $z_1$ bond in Fig. 2c) and the other bonds in the system play marginal role. In this case, at the ultimate tensile strength point, this aforementioned bond suddenly ruptures, which can be also concluded from the split of ELF contour (Fig. 3b) that was originally maximum at the bond center.

We next investigate the lattice thermal conductivity of NPG monolayer using the classical molecular dynamics simulations. Since the fixed boundary conditions employed in the NEMD may limit the contribution of some phonons, NEMD simulations we carried out for NPG samples with different lengths from ~50 nm to ~200 nm to consider the length effect. In Fig. 4, the NEMD results for the length effect on the room temperature lattice thermal conductivity of NPG along the armchair and zigzag directions are plotted. According to the results shown in Fig. 4, along the graphene nanoribbons (armchair direction) a general increasing trend in the predicted thermal conductivities is observable as the sample length increase. On the other hand along the zigzag direction, for the NPG lengths we initially considered (50-200 nm), only slight variations in the thermal conductivity values can be seen and therefore one can conclude that the thermal transfer is acceptably within the diffusive heat transfer. In this case, we conducted the NEMD simulation for the NPG monolayers with smaller lengths of 25 nm and 37 nm. As shown in Fig. 4 results, by including these additional points, the initial increasing trend in the thermal conductivity by increasing the NPG length also becomes obvious for the heat transfer along the zigzag direction. As a common approach, the phononic thermal conductivity of NPG monolayer with infinite length, $k_\infty$, can be estimated by the extrapolation of the NEMD results for the NPG samples with finite lengths, $k_L$ [80]. In this approach, the length dependence, $L$, of the thermal conductivity can be described by [80,81]:

$$\frac{1}{k_L} = \frac{1}{k_\infty}\left(1 + \frac{\Lambda}{L}\right) \qquad (1)$$

Here, $\Lambda$ is the effective phonon mean free path. As shown in Fig. 4 results by the continuous lines, we used the Eq. 1 for the fitting to the acquired NEMD data points. On this basis, the length independent room temperature phononic thermal conductivity of single-layer and free-standing NPG along the armchair and zigzag directions were calculated to be 14.1±0.8 W/m.K and 6.0±0.4 W/m.K, respectively. The anisotropic thermal transport along the NPG is clearly in accordance with our results concerning the mechanical properties. Such an anisotropicity in the thermal conductivity can be easily explained, by taking into account that in the worst cases, along the armchair direction (graphene nanoribbons) a factor of three more bonds can be simultaneously involved in the phonon transport as compared with the zigzag direction. It is worth noting that the predictions by the classical MD simulations strongly depend on the choice of interatomic potentials in describing the atomic interactions. We should therefore remind that by employing the similar extrapolating technique (Eq. 1), the thermal conductivity of pristine graphene using the AIREBO force-field was recently predicted to be 709.2 W/m.K [82]. As it is clear, NPG exhibits around two orders of magnitude suppressed thermal conductivity than pristine graphene. To facilitate the future studies, we included a super-cell of NPG monolayer in the data section which was fully relaxed by the MD simulations at the room temperature using the NPT method.

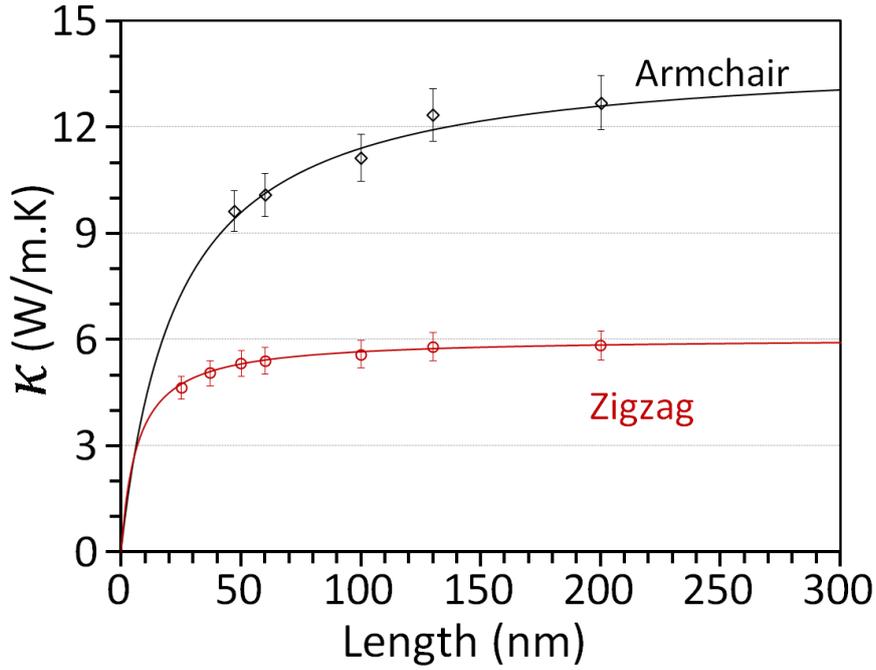

**Fig. 4**, Classical NEMD results for the length effect on the lattice thermal conductivity of single-layer nanoporous graphene along the armchair and zigzag directions calculated at 300 K. The continuous lines illustrated the fitted curves using the Eq. 1.

In order to explore the electronic properties of NPG monolayer, the band structure along the high symmetry X-Γ-Y directions and the total density of states (DOS) were calculated and the obtained results on the basis of GGA/PBE are illustrated in Fig. 5. Our results indicate that the GNP monolayer is a direct band-gap semiconductor at the Γ point (Fig. 5a) in agreement with previous reports [59,83]. We predicted a band-gap value of 0.68 eV, which is slightly lower than the previous theoretical predictions of 0.74 eV by Moreno *et al*. [59] and ~0.7 eV by Calogero *et al*. [83]. Worthy to remind that pristine graphene has a zero band-gap electronic character, with four half-filled degenerated states at the intrinsic Fermi level. The 4-fold degeneracy, consisting of two degenerated states at two nonequivalent Dirac points (K and K'), originates from the crystal symmetry of graphene. On this basis, band-gap opening in graphene can be achieved by breaking the symmetry in the pristine honeycomb lattice of graphene [84,85]. In the NPG nanosheets due to the breaking of the crystal symmetry, the $p_z$ orbitals of the C atoms, which are responsible for the DOS around the Dirac point in pristine graphene, are suppressed and consequently lead to a band-gap opening.

Since the PBE functional is well-known to underestimate the band-gap values of semiconductors, the electronic density of states was also computed using the HSE06 hybrid functional. Based on the acquired DOS by HSE06 method (see Fig. 5c), the band-gap of NPG monolayer was estimated to be 0.88 eV. It is worthy to remind that pristine graphene shows a semi-metallic character with a zero band-gap and the application of graphene in semiconductor devices requires a band-gap opening in order to switch the conductivity between on and off states. Therefore NPG nanosheets can serve as excellent alternatives for graphene to be used in semiconducting devices. In this work the band-gap was only calculated for the single-layer and stress free NPG nanosheet. Nevertheless, the effects of number of atomic layers (thickness) and strain on the band-gap of NPG can play critical roles in its practical application as a semiconductor, which should be investigated in separate studies.

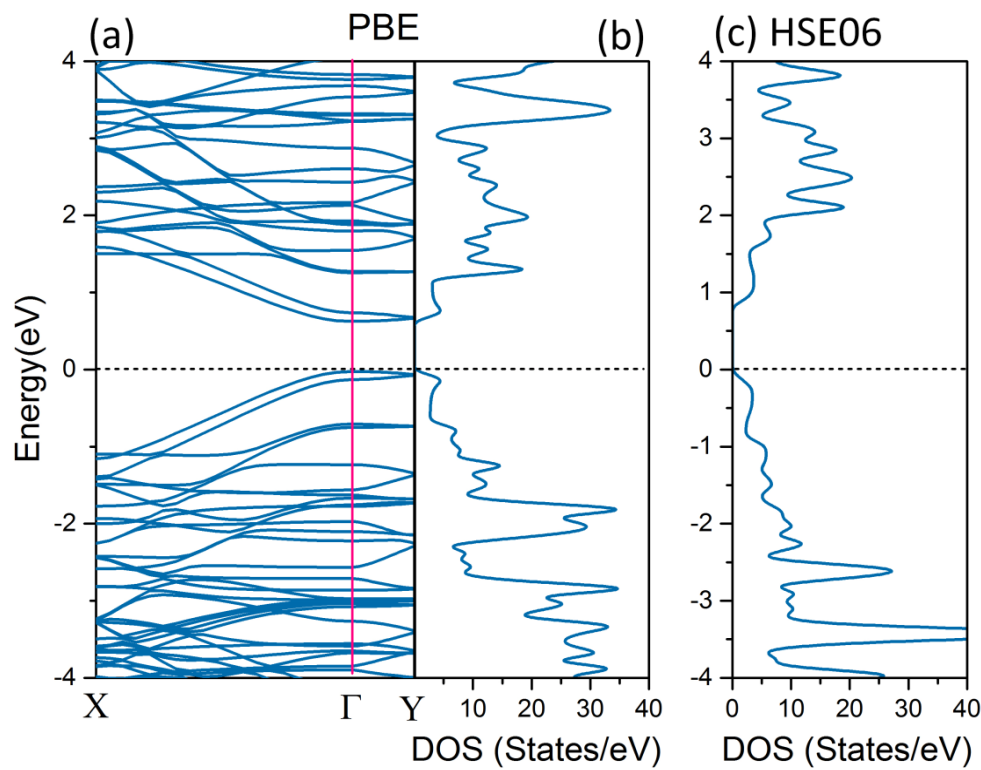

**Fig. 5**, Band structure and total density of states (DOS) of nanoporous graphene monolayer predicted by the PBE functional. The DOS is also plotted using the HSE06 method to more accurately report the band-gap. The Fermi energy is aligned to zero.

Since our electronic calculation results confirm the semiconducting electronic character of NPG monolayer, investigation of optical properties of this novel 2D material can be appealing for its practical applications in optoelectronics. In this case, we also compared the optical properties of NPG monolayer with those of the pristine graphene [69,71]. Because of the considerable depolarization effect in the 2D planar geometry for light polarization normal to the atomic-plane, we only focus on the optical absorption spectrum for the light polarization along the planar directions [86,87]. Moreover, the asymmetric atomic structure of NPG along the armchair and zigzag direction may also result in anisotropic optical spectra for the light polarizations along the armchair (E||arm) and zigzag (E||zig) directions. Hence, the optical properties for light polarizations along the both planar directions were calculated and reported. The imaginary and real parts of the dielectric function ($Im_{\alpha\beta}(\varepsilon)$ and $Re_{\alpha\beta}(\varepsilon)$, respectively) of NPG and graphene monolayers versus photon energy for the both in-plane directions obtained from RPA + PBE are illustrated in Fig. 6. The absorption edge of $Im_{\alpha\beta}(\varepsilon)$ of NPG occurs at 0.40 eV and 0.66 eV along E||arm and E||zig, respectively. In contrast, the $Im_{\alpha\beta}(\varepsilon)$ of graphene along armchair direction consists of a very significant peak at small frequencies and also another peak occurring at 4.1 eV, while along zigzag direction the absorption edge starts at 0.40 eV. For higher energies than 3.0 eV, the optical response of graphene overlaps in E||arm and E||zig. Values of static dielectric constant (value of dielectric function at zero energy) for NPG monolayer for E||arm was found to be 5.53, while in the case of E||zig it was found to be 3.09. The corresponding values for the pristine graphene for E||zig was 5.86 while the real part of the dielectric function shows a singularity at zero frequency along E||arm, thus exhibiting a metallic optical character.

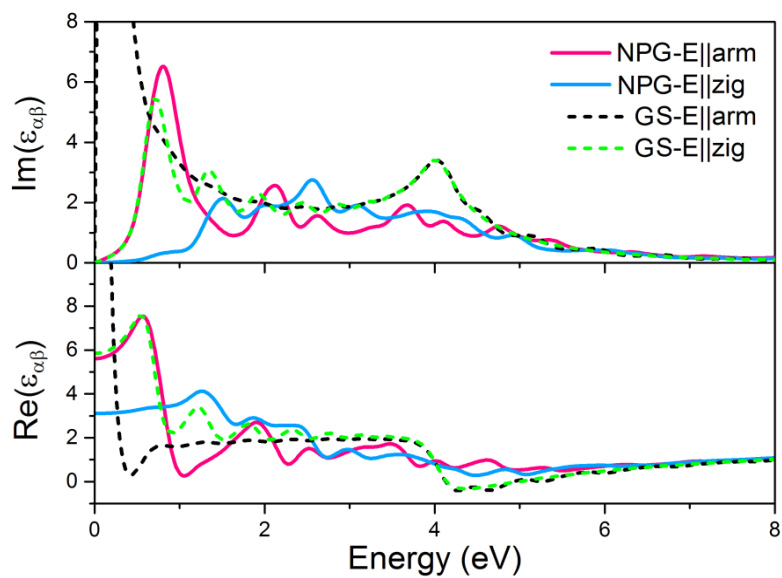

**Fig. 6**, Imaginary and real parts of the dielectric function of single-layer nanoporous graphene (NPG) and pristine graphene (GS) for the in-plane (parallel to the armchair-axis (E||arm) and parallel to the zigzag-axis (E||zig)) light polarizations, predicted using the PBE plus RPA approach.

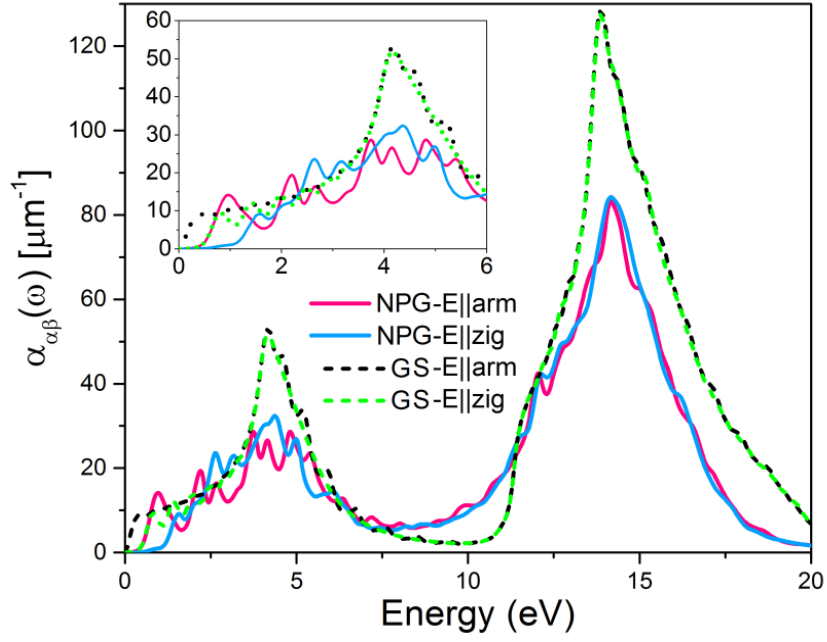

**Fig. 7**, Absorption coefficient of single-layer nanoporous graphene (NPG) and pristine graphene (GS) for the in-plane (parallel to the armchair-axis (E||arm) and parallel to the zigzag-axis (E||zig)) light polarizations, predicted using the PBE plus RPA approach. Insets show amplified regions of the absorption coefficient in the low frequency regime.

The absorption coefficients $\alpha_{\alpha\beta}$ for NPG monolayer are plotted in Fig. 7. In this case we also compared acquired results with the absorption coefficient of pristine graphene for the both planar polarization directions. The acquired results reveal that the first absorption peaks for NPG occur at 0.98 eV and 1.58 eV along E||arm and E||zig, respectively, which are in IR and near IR (NIR) range of light. The second absorption peaks for this system happen at 2.19 eV and 1.98 eV along E||arm and E||zig which are in the visible range of light. Our results of absorption coefficient show that this monolayer is highly desirable for the practical applications in optoelectronic devices operating in IR, NIR and visible spectral range. The first absorption peak of pristine graphene along E||arm occurs at 0.38 eV while it locates at 0.83 eV for E||zig. It is well obvious that the first absorption peaks of NGP experience a blue shift in comparison with graphene. In general, both NPG and graphene monolayers yield two main peaks in the absorption coefficient for the electric field parallel to the atomic layer. First main peaks for the both systems are sharp and occur around 4.00-5.00 eV, which is related to $\pi \rightarrow \pi^*$ transitions. Second main peak is broad and occurs around 14–15 eV, related to both $\pi \rightarrow \pi^*$ and $\sigma \rightarrow \sigma^*$ transitions. Some weak peaks are also observable between these two peaks, which are related to weak resonances. It is worthy to note that because of optical selection rules, only $\pi \rightarrow \pi^*$ and $\sigma \rightarrow \sigma^*$ transitions are allowed if the light is polarized parallel to the planar directions, in contrast, only $\pi \rightarrow \sigma^*$ and $\sigma \rightarrow \pi^*$ transitions are allowed for perpendicular polarization to graphene layers [88]. In this work we did not explore the effects of mechanical straining on the optical properties of NPG nanosheets, which is an attractive topic and should be investigated in the future studies.

**4. Concluding remarks**

In a latest experimental advance, a bottom-up synthesis method was successfully devised to fabricate nanoporous graphene. Nanoporous graphene (NPG) includes ordered arrays of nanopores separated by graphene nanoribbons. Motivated by this exciting experimental accomplishment, in this work we explored the mechanical response, thermal conductivity and electronic/optical responses of single-layer NPG using the first-principles density functional theory and classical molecular dynamics simulations. Our theoretical results clearly show highly anisotropic mechanical properties and thermal conductivity of NPG. Along the armchair direction (nanoribbons length) NPG yields higher tensile strength and thermal conductivity by factors of 1.6 and 2.3, respectively, as compared with the zigzag direction (nanoribbons width). This novel 2D carbon based material was found to exhibit higher rigidity, tensile strength, ductility and stretchability, along the armchair direction as compared with the zigzag direction. Notably, NPG nanosheet was predicted to yield high elastic modulus of 174 N/m and 144 N/m and remarkable tensile strengths of 15.0 N/m and 9.4 N/m along the armchair and zigzag directions, respectively. Interestingly, whilst the elastic modulus and tensile strengths of NPG are only by around three-folds lower than those of the graphene, it shows around two orders of magnitude suppressed thermal conductivity than graphene. According to non-equilibrium molecular dynamics results on the basis of AIREBO force-field, the room temperature lattice thermal conductivity of NPG nanosheets along the armchair and zigzag directions were estimated to be 14.1±0.8 W/m.K and 6.0±0.4 W/m.K, respectively. First-principles calculations within the GGA/PBE functional confirm that single-layer NPG shows direct band-gap semiconducting electronic characters at the Γ point, with a band-gap of 0.68 eV. According to HSE06 method, the band-gap of NPG monolayer was estimated to be 0.88 eV. The analysis of optical properties indicate that the first absorption peaks of NPG locate at 0.98 eV and 1.58 eV for the light polarization along armchair and zigzag directions, respectively, which are in IR and near IR (NIR) range of light, whereas the corresponding second absorption peaks position at 2.19 eV and 1.98 eV, respectively, in the visible range of light. The first absorption peaks reveal that NPG nanomembranes can absorb the visible, IR and NIR light, suggesting their prospect for the applications in optoelectronics and nanoelectronics. The acquired results by extensive theoretical simulations provide a comprehensive vision concerning the critical properties of this novel nanoporous carbon based 2D semiconductor and will hopefully serve as a guide for the future theoretical and experimental studies.

**Acknowledgment**

B. M. and T. R. greatly acknowledge the financial support by European Research Council for COMBAT project (Grant number 615132). B. M. and X. Z. particularly appreciate the funding by the Deutsche Forschungsgemeinschaft (DFG, German Research Foundation) under Germany's Excellence Strategy within the Cluster of Excellence PhoenixD (EXC 2122, Project ID 390833453).

**Data availability**

The energy minimized unit-cell by the DFT, the relaxed super-cell for the MD modelling, the script to load the model in LAMMPS are available to download from:
https://data.mendeley.com/datasets/8vywj9gw84/draft?a=d3f5d875-15ae-4a11-9aa2-9ed436b61d26